\documentstyle[mprocl,psfig]{article}

\def\Journal#1#2#3#4{{#1} {\bf #2}, #3 (#4)}

\def\sm{\mbox{M}_{\odot}}

\begin{document}

\title{MACHO versus HST: how bright can dark matter be?}

\author{E. J. Kerins}

\address{Observatoire Astronomique de Strasbourg, 11 Rue de
l'Universit\'e, F-67000 Strasbourg, France.}


\maketitle\abstracts{Recent results from gravitational microlensing
experiments, such as MACHO, indicate that a substantial fraction of
the Galactic dark matter (DM) is in compact form, with a typical mass
in the range $\sim 0.05 - 1~\sm$. This mass range favours the DM being
either low-mass hydrogen-burning stars or white-dwarf remnants of an
early stellar population. There appears however to be a conflict
between microlensing and other results, such as from Hubble Space
Telescope (HST), which argue against a high DM fraction in
hydrogen-burning stars or their remnants. Here I discuss the
possibility that the DM detected by MACHO comprises low-mass stars
residing in globular cluster associations. I show that such a scenario
can reconcile HST and MACHO results and can also satisfy cluster
dynamical constraints.}
  
\section{Results from MACHO and HST}

Gravitational microlensing results\cite{jet} towards the Large
Magellanic Cloud obtained in the first 2 years of the MACHO
experiment\cite{alc97} indicate that $20-80\%$ of the DM comprises
objects with masses between $0.05 - 1~\sm$. These statistical
estimates are highly model dependent, though the quoted spread in
values is typical of a number of near-isotropic halo
models\cite{eva94} considered by the MACHO collaboration\cite{alc97}.

In principle, the lenses responsible for the MACHO results need not be
baryonic but may instead be in some other form of compact matter, such
as primordial black holes. However, primordial black holes must form
at a relatively late epoch in the early universe in order to explain
the microlensing mass scales, in which case they may modify
nucleosynthetic predictions for the production of the light
elements. Such a modification would almost certainly destroy the very
precise correspondence between the predicted and observed
light-element abundances, a correspondence which forms one of the
cornerstones of Big-Bang cosmology.

Assuming the lenses are baryonic in origin, the MACHO mass estimates
implicate low-mass stars or white-dwarf remnants. Sub-hydrogen-burning
``stars'', or brown dwarfs, are not quite excluded, though their
allowed mass is confined to a rather narrow range. However, analyses
of HST fields obtained in the $V$ and $I$ bands place strong limits on
the halo contribution of both low-mass stars and white dwarfs. White
dwarfs are limited to a halo fraction below $10\%$ if their age does
not exceed 14~Gyr\cite{gra97}, and low-mass stars must contribute no
more than $4\%$ to the DM if they have solar metallicity\cite{bah94},
or less than $1\%$ if their metallicity is the same as or less than
that of Population~II stars\cite{gra96,ker97a,ker97b}. Thus, it
appears difficult to reconcile the microlensing and HST results.

\section{Constraints on clusters}

In this talk I choose to re-examine the HST constraints on low-mass
stars. In obtaining the above constraints one makes the very strong
assumption that the local dark halo stellar distribution is perfectly
smooth on scales comparable to the observation volume. How do such
limits change if the stellar distribution is not smooth but instead
clumpy?  Specifically, consider a scenario in which the halo DM
comprises dark globular clusters of very low-mass (VLM) stars with
mass $m$ just above the hydrogen-burning limit. If these stars have a
more or less primordial composition then they are effectively metal
free, in which case the condition for hydrogen burning is $m > m_{\rm
burn} = 0.092~\sm$\cite{sau94}.

The assumption of clustering is motivated both by the observation of
visible globular cluster systems in the inner Galaxy and by
theoretical DM formation scenarios\cite{ash90,dep95} which explain the
existence and distribution of visible globular clusters and predict
the existence of a population of dark globular clusters. Such dark
clusters are predicted to have a more or less isothermal
distribution (as required to explain the halo DM problem) and comprise
either brown dwarfs or VLM stars.

The additional assumption of clustering introduces an additional form
of constraint, namely dynamical constraints on populations of very
massive halo objects\cite{car94}. Such constraints place lower and
upper bounds on both the cluster mass $M$ and radius $R$ for a given
halo fraction $f_{\rm h}$. The constraints come from a variety of
considerations: tidal forces acting across the clusters;
cluster--cluster collisions; evaporation timescales; and their
disruptive effects on visible cluster populations.

\begin{figure}
\psfig{figure=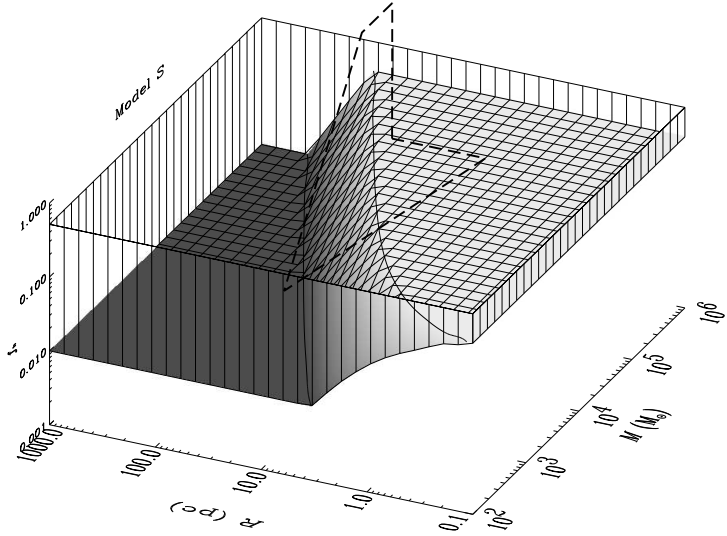,height=3.0in,angle=90}
\caption{MACHO, HST and dynamical constraints on VLM star clusters of
mass $M$ and radius $R$ as a function of allowed halo fraction
$f_{\rm h}$. The upper plateau to the right shows the
MACHO $95\%$ {\em lower}\/ limit on $f_{\rm h}$ from optical depth
estimates. The lower plateau to the left shows the HST $95\%$ {\em upper}\/
limit on VLM stars in the limiting case of a smooth stellar
distribution. The curved surface adjoining the two plateau is the HST
$95\%$ {\em upper}\/ limit on clusters. This limit continues to rise
asymptotically above the MACHO plateau but is truncated here for
clarity. The dashed lines projected onto the MACHO lower-limit plane
bound cluster parameters which satisfy all dynamical constraints. Its
intersection with the MACHO plateau indicates cluster parameters which
satisfy MACHO, HST and dynamical constraints. A star mass of $m =
0.092~\sm$ is assumed.
\label{f1}}
\end{figure}

Figure~\ref{f1} brings together the MACHO, HST and dynamical
constraints for dark clusters comprising zero-metallicity VLM stars
with mass $m = m_{\rm burn}$\cite{ker97b}. The HST limits are derived
from an analyses of 51 fields obtained by Gould {et
al}.\cite{gou97} The assumed halo model is a simple cored isothermal
sphere, denoted model~S by MACHO in its halo model
analysis\cite{alc97}, and has a local density $\rho_0 =
0.008~\sm$~pc$^{-3}$. Evidently, clusters with mass in the range $M
\sim 10^3 - 10^6 ~\sm$ and radius $R \sim 3-100$~pc can satisfy all
dynamical constraints whilst reconciling the MACHO and HST results.

The one caveat for the model is that one requires a very large
fraction of the stars, more than $95\%$, to remain inside the clusters
until the present. This is because the HST observations limit the
fraction of stars which can have escaped from the clusters to
form a smooth background. However, if the escaped stars have not
yet phase mixed sufficiently to form a smooth background then the
cluster membership constraint can be weakened\cite{ker97b}.

\section{Summary}

Whilst HST appears to rule out compact baryonic objects in the mass
interval $0.1 - 1~\sm$ from comprising more than $\sim 10\%$ of the
halo DM, microlensing results indicate that such objects provide
between $20-80\%$ of the DM. 

A scenario in which a substantial fraction of the halo DM comprises
dark globular clusters of VLM stars can explain the apparently
contradictory results from MACHO and HST, whilst satisfying all known
constraints arising from their dynamical effects.

\section*{Acknowledgements}
This work is supported by the EU through a Marie Curie TMR
Postdoctoral Fellowship.

\end{document}